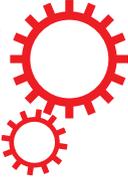

**OPEN**

# Strong coupling of magnons in a YIG sphere to photons in a planar superconducting resonator in the quantum limit

R. G. E. Morris, A. F. van Loo, S. Kosen & A. D. Karenowska

We report measurements made at millikelvin temperatures of a superconducting coplanar waveguide resonator (CPWR) coupled to a sphere of yttrium-iron garnet. Systems hybridising collective spin excitations with microwave photons have recently attracted interest for their potential quantum information applications. In this experiment the non-uniform microwave field of the CPWR allows coupling to be achieved to many different magnon modes in the sphere. Calculations of the relative coupling strength of different mode families in the sphere to the CPWR are used to successfully identify the magnon modes and their frequencies. The measurements are extended to the quantum limit by reducing the drive power until, on average, less than one photon is present in the CPWR. Investigating the time-dependent response of the system to square pulses, oscillations in the output signal at the mode splitting frequency are observed. These results demonstrate the feasibility of future experiments combining magnonic elements with planar superconducting quantum devices.

Magnons are elementary excitations of magnetically ordered materials. For more than half a century, their study has been recognized as a fertile area of research; both for its rich fundamental physics, and its potential technological applications. Experimentally, the magnon modes that can be accessed in a given magnetic sample depend on its geometry and magnetic environment[1,2].

While magnons can be studied in a wide range of magnetic materials, one in particular stands out for its remarkable properties. Yttrium-iron garnet (YIG) is a ferrimagnetic garnet with extremely low magnon damping[3]. This low loss, in combination with very low electrical conductivity, makes it ideal for the production of high-$Q$ microwave resonators and waveguides. Magnon systems based on YIG have been used for some decades as the basis for commercial radio-frequency devices and components[4–6]. Much more recently, the use of such systems in the development of devices for quantum computation has begun to attract significant attention[7]. This has led to a number of investigations into architectures for coupling magnons to photons in ways that can easily be extended into the quantum regime, in the new field of quantum magnonics.

To date, work towards quantum magnonics includes room-temperature[8–11] investigations into the coupling of standing magnon modes to photons in 3D electromagnetic cavities, as well as several experiments performed at low temperatures approaching the quantum regime[12–16]. In addition, the coupling of a superconducting qubit to magnons via a 3D cavity[17] has been demonstrated. However, there has been little work done with planar superconducting structures[18], perhaps due to the difficulty of keeping films superconducting in the magnetic fields required for magnon excitation. Quantum devices built using these 2D systems offer the ability to use qubit-specific control lines to precisely control qubit states[19], typically have high qubit-resonator coupling, and are well-suited to integration with other electronic components[20]. As well as opening up the possibility for novel hybrid devices incorporating magnonic systems, coupling such structures to YIG would offer new tools to study the quantum physics of magnons.

In this paper we investigate the coupling of magnons in a YIG sphere to photons in a coplanar waveguide resonator (CPWR). Our experimental geometry allows us access to a variety of magnon modes with different characteristic frequencies, and to identify some of these modes based on numerical calculations. We also investigate the time-dependent behaviour of the system and extend our measurements into the quantum regime by driving

Clarendon Laboratory, Department of Physics, University of Oxford, Oxford, UK. Correspondence and requests for materials should be addressed to R.G.E.M. (email: richard.morris@physics.ox.ac.uk)

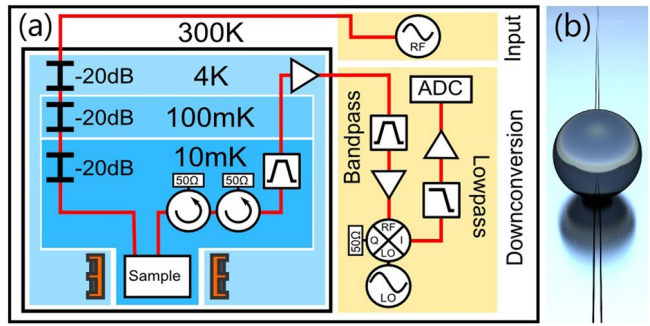

**Figure 1.** Illustrations of the experimental apparatus used. (**a**) The sample is mounted inside a dilution refrigerator where a superconducting magnet provides a magnetic field, allowing tuning of the magnon mode frequencies. Coaxial transmission lines connect the input to a microwave source, and the output to downconversion instrumentation. Approximately 70 dB of attenuation is present in the input line (60 dB from fixed attenuators and 10 dB from the coaxial cables). (**b**) Schematic showing the scale of the YIG sphere, with diameter 250 µm, compared to the CPWR, with centre conductor width 10 µm.

the CPWR such that it contains, on average, less than one excitation. This work validates our understanding of the behaviour and interactions of magnon systems at millikelvin temperatures, and is a necessary step in the development of experimental systems coupling magnons to 2D circuit quantum electrodynamic (cQED) systems.

## Methods

Our experiments are performed using a half-wavelength CPWR fabricated from a 150 nm film of superconducting niobium. The CPWR has a frequency of 4.572 GHz and a quality factor of 1220 at zero applied magnetic field. A monocrystalline YIG sphere of diameter 0.25 mm, obtained from Ferrisphere Inc.[21], is glued onto the centre of the CPWR where the amplitude of its magnetic field is maximal. The sample is mounted inside an oxygen-free copper box which is thermally anchored to the mixing chamber plate of a dilution refrigerator with a base temperature of approximately 10 mK. This low temperature is necessary to ensure that there is a negligible thermal population of excitations in the frequency range we investigate. The position of the sample is such that it sits at the centre of a superconducting magnet providing a field parallel to the length of the CPWR and in the plane of the niobium film. A schematic of the experimental setup and the room temperature microwave apparatus is shown in Fig. 1a. We are also able to measure the phase of the transmitted signal by comparing the output signal from the sample with a downconverted reference signal from the microwave source (not shown in Fig. 1).

Experiments are performed by measuring the complex transmission of the CPWR as a function of frequency, magnetic field and input signal power. Approximately 70 dB of attenuation between the output of the room-temperature microwave source and the input port of the CPWR ensures that the electrical noise temperature of the input signals is comparable to the thermodynamic temperature of the sample. From the output of the CPWR, signals are amplified by approximately 40 dB at 4 K before reaching the room temperature part of the measurement instrumentation where, after mixing to an intermediate frequency and undergoing further amplification and filtering, they are digitized at 2.5 GHz using a fast data acquisition card. Measurements were typically averaged between ten thousand and one hundred thousand times.

The frequencies of the magnon modes in the YIG sphere are functions of the magnetic bias field, and can therefore be shifted by adjusting the current in the superconducting magnet. Since the plane of the CPWR is aligned parallel to the applied magnetic field, it remains superconducting up to relatively high fields, making it possible to bring a variety of magnon modes into resonance with it. By calculating the shape of the resonance microwave magnetic field of the CPWR and the magnetisation within the YIG sphere for a given magnon mode, we can find their relative orientations and thus estimate the relative strength of the coupling for different magnon modes.

All the data displayed in this study, as well as details of the calculations performed, can be obtained from the corresponding author on request.

## Results

**Theoretical calculations.** The microwave field of the CPWR in the region of the sphere was computed using HFSS[22] (see Fig. 2a). In contrast with the experimental geometries used in some of the previous investigations on this topic[8, 11–13, 17], the magnetic field of the CPWR is strongly inhomogeneous in the region around the centre conductor, making it highly non-uniform in the region of the YIG sphere. This allows us to address a significant number of different magnon modes in the sphere.

The magnon modes in ferromagnetic spheres are traditionally described using three indices $(n, m, r)$[1], where $n$ and $m$ refer to the order of Legendre polynomials used in determining the resonance condition, and $r$ distinguishes between modes when there are multiple solutions for the same values of $n$ and $m$. The mode frequencies and the associated distribution of magnetisation transverse to the magnetic field can be calculated from the theory presented in ref. 23. None of the modes relevant to this experiment have multiple solutions, so in what follows modes will simply be identified by their $n$ and $m$ indices. The relative coupling strength of different magnon modes can be estimated from the dot product of the magnetisation with the CPWR magnetic field over the sphere, as illustrated for the (2, 2) magnon mode in Fig. 2b.

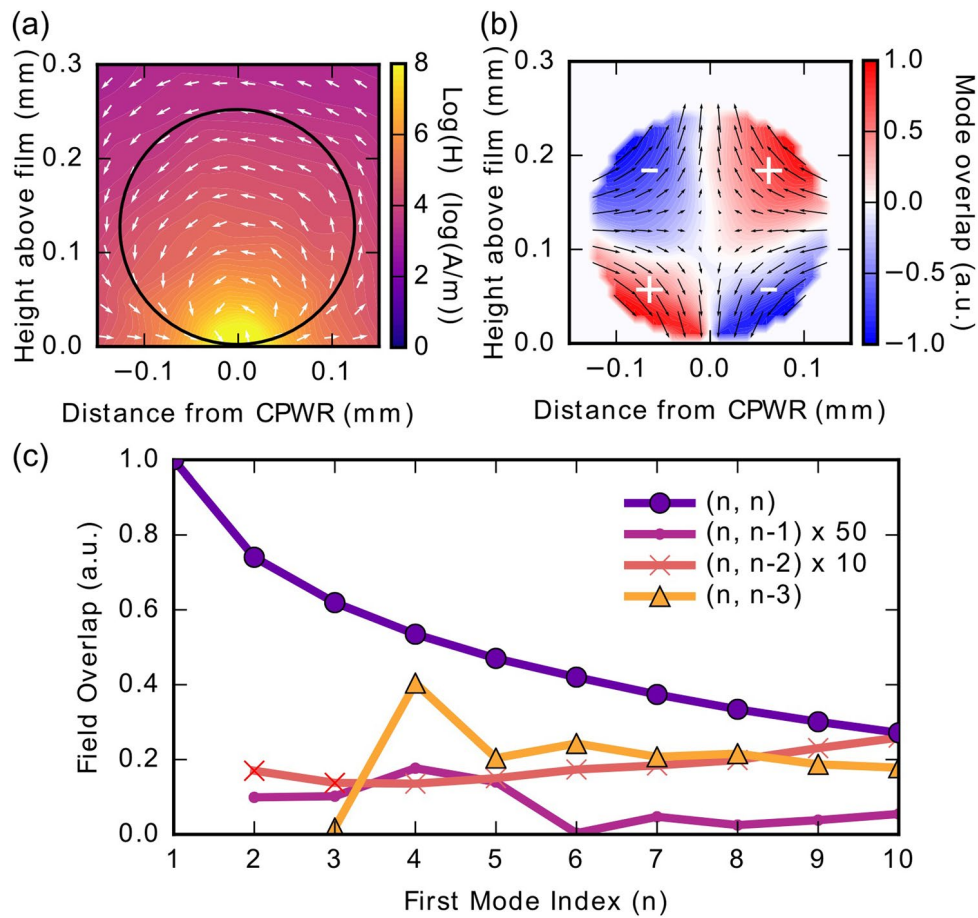

**Figure 2.** Results of the theoretical analysis relating to the magnon modes in the YIG sphere. (**a**) The microwave magnetic field distribution above the centre of the CPWR. A circle of equivalent size to the YIG sphere used in the experiment is superimposed. As can be seen, the field inside the sphere is highly non-uniform. (**b**) The overlap of the CPWR magnetic field with the (2, 2) magnon mode in the sphere. The arrows indicate the pattern of magnetisation in the sphere for this mode, and the colour scale reflects the dot product of the CPWR field and sphere magnetisation at each point. (**c**) The relative coupling of four families of sphere modes with the CPWR field is estimated for values of $n$ up to 10. Lines are guides to the eye. The $(n, n-1)$ and $(n, n-2)$ families are scaled as indicated to make them more visible. The $(n, n)$ modes consistently couple most strongly, which remains the case for small displacements of the sphere from the centre of the CPWR.

The results of relative coupling calculations for magnon modes in three families up to $n=10$ are summarised in Fig. 2c. The $(n, n)$ modes consistently couple most strongly to the CPWR. The (1, 1) mode, also known as the Kittel or ferromagnetic resonance mode, has the strongest coupling of all. This mode corresponds to in-phase precession of all spins in the sphere.

**Experimental measurements.** To observe the coupling between the YIG sphere and the CPWR experimentally, we measure the complex transmission (S21) of microwave radiation through the CPWR as a function of frequency and magnetic field. The results of this experiment are summarised in Fig. 3. Each data point in the 2D plots is averaged 4000 times. The results are consistent over a range of temperatures from 10 mK to 5 K, and are reproducible in measurements made over a period of several months.

As the field increases above 120 mT, photons in the CPWR couple to different magnon modes in the YIG sphere. This is observed as a series of avoided crossings, with the Kittel mode, labelled (1, 1) in Fig. 3a, being the most pronounced. Apart from the Kittel mode, at least ten additional splittings are visible. As discussed above, we identify the strongly coupled modes by considering the magnetisation distribution and fit their frequencies using the $(n, n)$ mode family, which converges at high $n$, as seen in the data. We also observe some weakly coupled modes of similar slope attributable to other mode families in the sphere, such as those plotted in Fig. 2c.

### Discussion

There is good agreement between the calculated frequencies of the $(n, n)$ modes and our experimental data up to the (4, 4) mode, for a saturation magnetisation of 170 kA/m. For higher modes, the observed frequencies begin to diverge from the predicted frequencies, which we attribute to the effect of magnetocrystalline anisotropy. This effect is likely to be especially significant at low temperatures as the first-order anisotropy constant of YIG is approximately three times larger near zero Kelvin than at room temperature[24].

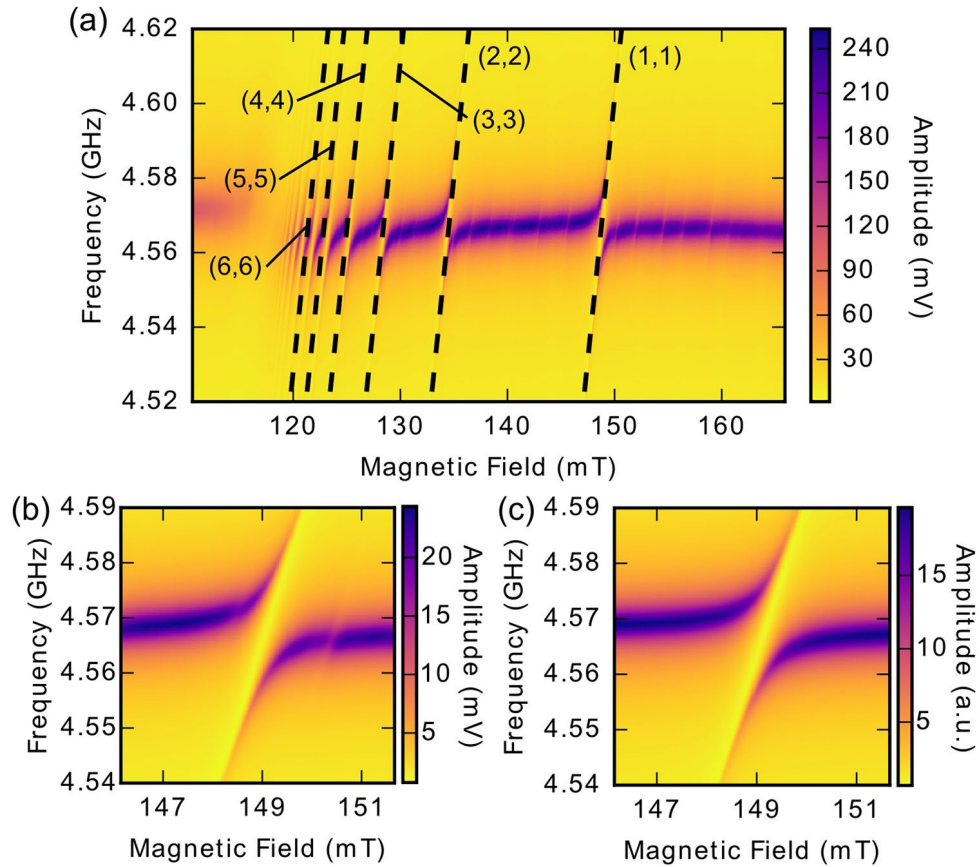

**Figure 3.** Results of microwave spectroscopy measurements of the sample. (**a**) Experimental transmission through the CPWR as a function of frequency and applied magnetic field at an input power of approximately −80 dBm. More than ten anticrossings are visible, along with a number of weaker features. The (n, n) modes are plotted using the equations in ref. 23 with a saturation magnetisation of 170 kA/m and a constant offset to account for the effects of magnetocrystalline anisotropy. (**b**) A high-resolution measurement of the (1, 1) anticrossing. The splitting is 16.34 MHz and two of the weaker features are also visible at higher and lower magnetic field. (**c**) A fit to the crossing in panel (b) using Eq. 1. The two weaker crossings are not modelled.

If the bias magnetic field is along either the hard or easy axis of the crystal, the effect of anisotropy is to shift the frequencies of all modes in the spectrum by a constant factor[25]. However, in the case of a general orientation, the effects become much more difficult to calculate and analytic expressions for the magnon mode frequencies exist for only the first few low-order modes[26]. In our experiment, it was not practical to determine and align the orientation of the crystal axes of the sphere, making it impossible to calculate the exact frequencies taking anisotropy into account. The fitted lines in Fig. 3a use only a constant offset to account for anisotropy, which works well for the first few modes. The challenge of determining the effect of the anisotropy on higher-order mode frequencies also makes it difficult to identify the weakly coupled modes observed in the data with particular mode families.

The value of 170 kA/m used for the saturation magnetisation is significantly below the literature value of 197 kA/m at 4.2 K[27]. Our value is derived from the spacing of the (1, 1) and (2, 2) modes, where analytic expressions for the effects of anisotropy do exist. We were therefore able to calculate that anisotropy effects are not sufficient to explain this spacing, necessitating the use of a lower value of saturation magnetisation in our calculations.

Figure 3b shows a high-resolution measurement of the avoided crossing between the CPWR resonance and the Kittel mode. This crossing is fitted in Fig. 3c using an equation derived from the input-output formalism[12]:

$$S_{21}(\omega) = \frac{\kappa_{ext}}{i(\omega - \omega_r) - \kappa_{ext} - \frac{\kappa_{int}}{2} + \frac{g_m^2}{i(\omega - \omega_m) - \frac{\gamma_m}{2}}} \quad (1)$$

In this fit we account for the effect of anisotropy on the frequency of the Kittel mode $\omega_m$ since an analytic expression exists. Both amplitude and phase measurements were fitted simultaneously (phase information is not plotted in Fig. 3), resulting in a linewidth of $\gamma_m/2\pi = 2.97$ MHz for the Kittel mode, and internal and external linewidths of $\kappa_{int}/2\pi = 1.39$ MHz and $\kappa_{ext}/2\pi = 1.34$ MHz respectively for the CPWR. The average residual of the fit is below 4%, and deviations are only significant around the weaker features above and below the FMR field, which are not accounted for by Eq. 1. The coupling strength $g_m/2\pi = 8.17$ MHz is larger than either of the individual linewidths, indicating that the system is in the strong coupling regime. At these magnetic fields, the CPWR frequency is shifted down slightly from its zero-field value to $\omega_r = 4.568$ GHz. The trend in measured mode coupling strengths mirrors that predicted in Fig. 2c.

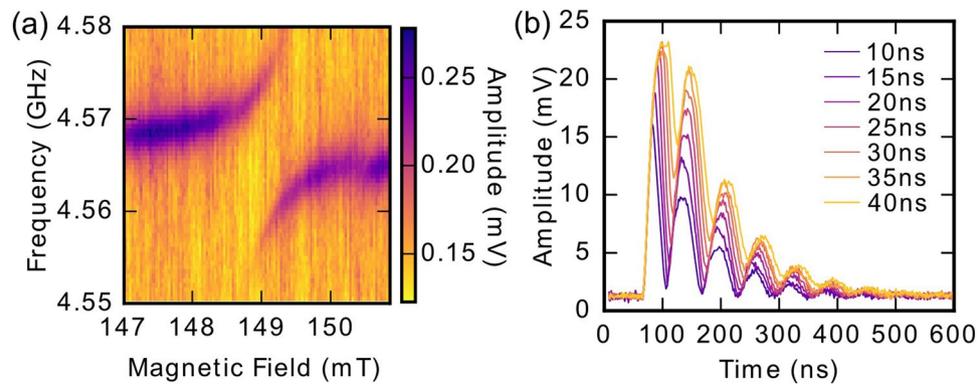

**Figure 4.** (**a**) Measurement in the quantum limit of the avoided crossing between the Kittel mode and the CPWR. The input power to the resonator is approximately −140 dBm, corresponding to, on average, less than one photon in the resonator. (**b**) Time response of the system to a square pulse of varying length when the applied magnetic field is tuned to bring the YIG sphere and the CPWR into resonance. Beating in the output signal is at the frequency of the level splitting.

By reducing the input power, the average number of photons in the CPWR can be reduced to less than one; the so-called quantum limit. Even at these low powers, high levels of averaging ($7.5 \times 10^5$ averages per data point) allow us to detect strong coupling between CPWR photons and magnon modes (see Fig. 4a), demonstrating the viability of the system in the context of potential quantum information processing applications.

We can also excite the YIG-CPWR system with a short square pulse having a carrier frequency between the two split levels (e.g. 4.570 GHz at 149 mT). After such a pulse, we see oscillations in the output signal with frequency equal to the splitting. These oscillations decay with an exponential envelope that is due to a combination of the intrinsic damping of the CPWR and the damping of magnons in the YIG sphere. The temporally narrow square pulse has a sufficiently wide frequency spectrum to simultaneously excite both levels, so that these oscillations are interpreted as being due to beating between the frequencies of the two levels. This interpretation is reinforced by the reduction in amplitude of the oscillations as the pulse length increases, reducing its spectral width.

In summary, we have demonstrated strong coupling of a magnonic resonator, consisting of a YIG sphere, to a coplanar waveguide resonator typical of those used in cQED experiments. The coupling strength of 8.17 MHz for the Kittel mode is in the strong coupling regime. We have also shown strong coupling to a series of other non-uniform modes of oscillation in the YIG sphere, which were identified as the $(n, n)$ modes.

The range and tunability of excitations available in magnonic systems presents interesting possibilities for building hybrid quantum systems. Our work differs from previous work in the chosen geometries of the magnonic and photonic resonators. While the work by Huebl et al.[18] did previously investigate the coupling of a CPWR to a magnonic system, that work focused on the FMR mode in a slab of gallium-doped YIG. Here, we show that the highly inhomogeneous field close to the CPWR allows for coupling to a variety of modes with different coupling strengths. The low magnon damping in our experiment also allows us the fully resolve the anticrossing, in contrast to the previous work.

Planar superconducting structures differ in various important ways from three-dimensional cavities for use in quantum information applications, as described in the introduction. By using superconducting qubits designed to work in sufficiently high magnetic fields[28], it is possible to integrate magnonic elements with planar quantum information devices. Our experiment therefore demonstrates the possibility of building a different class of hybrid quantum devices than those previously investigated[12,17].

### Acknowledgements

This work was supported by Engineering and Physical Sciences Research Council grant EP/K032690/1. S.K. would also like to thank the Indonesia Endowment Fund for Education for its support.

### Author Contributions

A.K. conceived the experiment; R.M., A.v.L., and S.K. carried out the measurements; R.M. and S.K. performed the calculations; R.M. prepared the manuscript; all the authors reviewed the manuscript.

### Additional Information

**Competing Interests:** The authors declare that they have no competing interests.

**Publisher's note:** Springer Nature remains neutral with regard to jurisdictional claims in published maps and institutional affiliations.